\documentclass[manuscript,screen]{acmart}

\begin{document}

\title{Data Acquisition Through Participatory Design for Automated Rehabilitation Assessment}

\author{Tamim Ahmed}
\authornotemark[1]
\email{tamimahm@usc.edu}
\affiliation{%
  \institution{Biomedical Engineering, University of Southern California}
  \city{Los Angeles}
  \state{California}
  \country{USA}
}
\author{Zhaoyi Guo}
\authornotemark[1]
\email{zhaoyigu@usc.edu }
\affiliation{%
  \institution{Computer Science, University of Southern California}
  \city{Los Angeles}
  \state{California}
  \country{USA}
}

\author{Mohammod Shaikh Sadid Khan}
\authornotemark[1]
\email{bit0328@iit.du.ac.bd}
\affiliation{%
  \institution{Institute of Information Technology, University of Dhaka}
  \city{Dhaka}
  \country{Bangladesh}
}

\author{Thanassis Rikakis}
\email{rikakis@usc.edu}
\affiliation{%
 \institution{Biomdeical Engineering, University of Southern California}
  \city{Los Angeles}
  \state{California}
  \country{USA}
  }
\author{Aisling Kelliher}
\email{akelliher@cinema.usc.edu}
\affiliation{%
 \institution{Media Arts + Practice Division, University of Southern California}
  \city{Los Angeles}
  \state{California}
  \country{USA}
  }

\keywords{Segmentation, Rehabilitation, ARAT, Rating, Calibration, Capture}


\maketitle

\section{Introduction}
The global aging population is driving a growing demand for effective and accessible physical rehabilitation services, particularly for conditions like stroke, where six months post-stroke, 60\% of survivors experience upper limb disability, with 40\% never regaining functionality \cite{kerimov2021effects}. To address this challenge, technology-assisted rehabilitation techniques are gaining popularity. These include virtual and augmented reality systems \cite{elor2018project, ramirez2014design} for motivation, robot-assisted therapy for high-intensity training \cite{morone2020robot}, and wearable sensors for objective feedback on daily activity \cite{kim2022use}. Home-based telerehabilitation and automated assessment of rehabilitation outcomes in the clinic can reduce the time clinicians spend on assessment and/or repetitive daily exercises. This allows clinicians to focus more on structuring and customizing rehabilitation. All technology-assisted rehabilitation systems rely partially on automated movement assessment.  Automated assessments must be compatible with clinician assessments so that clinicians can easily understand and use the automated results. 

In our previous research \cite{kelliher2020towards,ahmed2021automated}, we focused on automated assessment that is compatible with the clinical practice process by making clinicians' tacit knowledge explicit and computable. Clinicians typically use a hierarchical approach to evaluate rehabilitation tasks \cite{ahmed2021automated}. They assess the overall task but also break the performance of the task into segments and assess the segments as well as key movement components for each segment. Even though clinician methods for implementing hierarchical assessment vary, there is clear agreement on the need to assess the task, its segments, and movement quality components in an integrative manner. To make this approach effective, different segments and movement components require varied viewpoints for evaluation. For instance, to assess the initiation of movement of the upper extremity in stroke rehabilitation clinicians usually choose an ipsilateral viewpoint of the affected limb. After initiation of movement, they switch to a contralateral view so as to monitor finger orientation and hand grip during object engagement. Thus, we've found that the successful implementation of a semi-automated system involves multi-view data capture, standardized segmentation across different tasks, and expert rating of captured videos through a user-friendly interface that allows different clinicians to rate tasks, segments, and components in an integrative and standardized manner. 

Our previously implemented version of this capture-segment-rate workflow \cite{kelliher2020towards,ahmed2021automated,clark2021hybrid} involved a two-camera setup for capturing 15 tasks for upper extremity stroke rehabilitation. The selected tasks mapped well to key activities of daily living. We captured 9 stroke survivors performing multiple versions of these tasks. Four expert clinicians used the captures to establish a common segmentation vocabulary across the 15 tasks and associated movement quality components for each type of segment. We also designed a rating tool for therapists to rate tasks, segments, and movement components using a standardized rubric developed by the clinicians. In previous publications, we reported on key limitations of our approach that compromised the quality of the captured data and the rating data and thus impeded the use of the data for robust automated assessment.  Our two-camera system could not provide all the necessary viewing angles for a robust rating. We had research assistants hand-segment the videos which resulted in significant variations in the length of segments and significant variations in the chosen transition frame between segments. The capture and segmentation limitations resulted in clinicians not being able to consistently rate some of the captured videos. Furthermore, these limitations impeded our ability to apply our system to the automated implementation of standardized clinical assessment instruments such as the Action Research Arm Test (ARAT) for stroke rehabilitation \cite{mcdonnell2008action}.

In this paper, we report on three primary improvements to our system that have allowed us to overcome previous limitations and acquire high-quality data for developing algorithms for the automated assessment of the ARAT. Collaborating closely with rehabilitation experts we have successfully:\\
i) Engineered a bespoke four-camera setup with a simple user interface to enhance data acquisition capabilities. Clinicians were able to use the setup independently to capture 106 ARAT sessions with stroke survivors in the clinic. The setup can be generalized to unobtrusive capturing of upper extremity functional activity in any physical or occupational rehabilitation context.\\
ii) Developed a custom segmentation tool that was used by research assistants for standardized segmentation of the ARAT exercise videos. The segmentation tool can be generalized to standardized segmentation of videos of other instances of highly structured human movement (arts, sports, surgery, military training).\\
iii) Building upon our prior work on rating interfaces \cite{kelliher2020towards}, we also developed a rating tool for the clinicians to rate the segmented ARAT videos in a standardized manner so that the rated data can be used for the development of algorithms for the automated assessment of the ARAT. 

In the following sections, we describe the development, testing, and results of the implementation of these tools. We close the paper by proposing that the participatory design process used by our team to develop these tools (and acquire, segment, and rate the data for the ARAT) can be used in many other contexts where expert knowledge in human movement assessment needs to be integrated into the development of AI tools so as to make such tools compatible with expert practice and lead to augmented intelligence practice.   
\begin{figure}[!ht]
  \centering
  \includegraphics[width=0.7\linewidth]{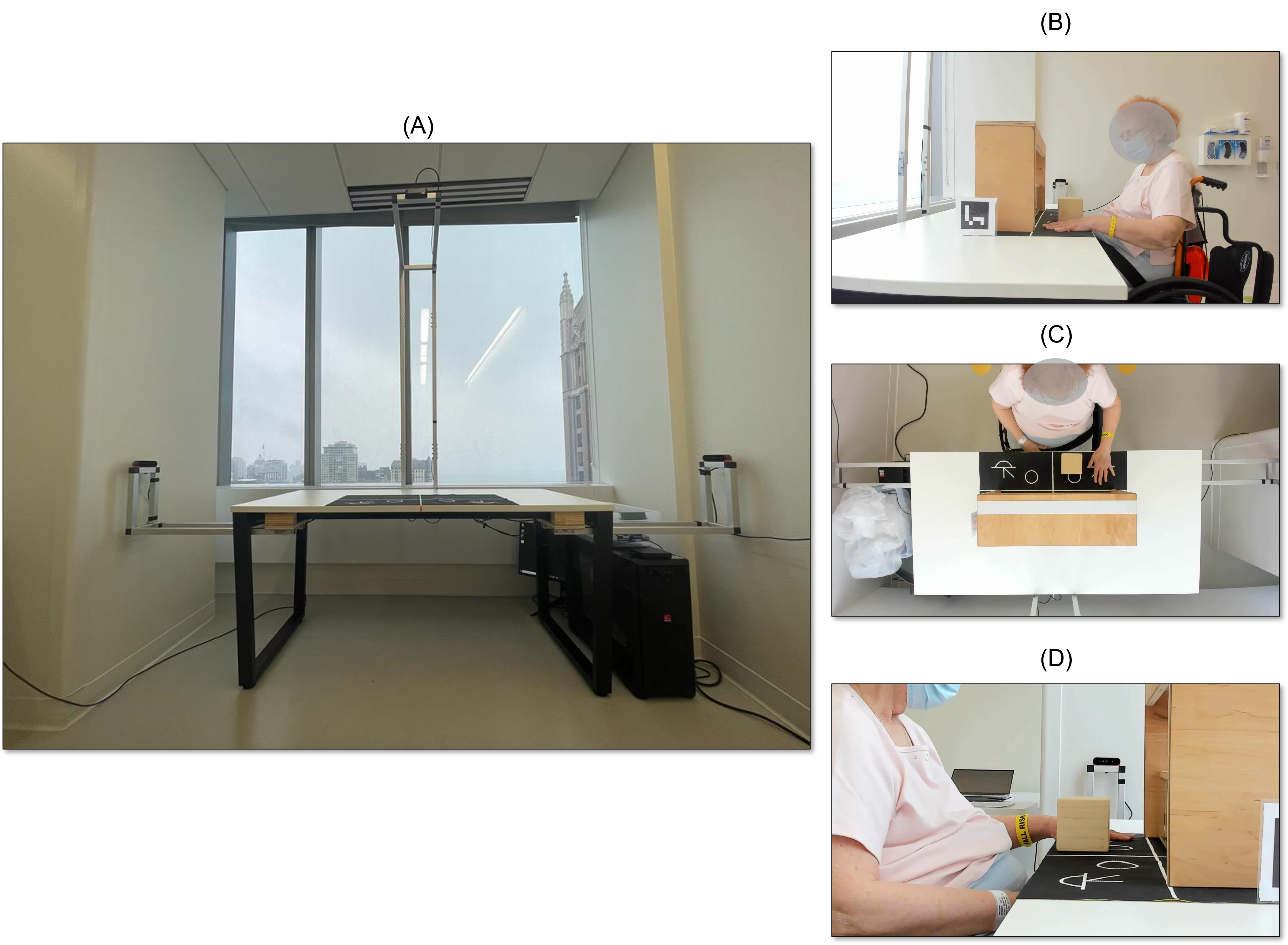}
  \caption{(A) The multi-camera ARAT exercise capture setup with the (B) ipsilateral, (C) transverse, and (D) contralateral view.}
  \label{fig:cap}
\end{figure}

\section{Methods}
\subsection{Standardized Capturing Setup for ARAT Exercises}
In ARAT, stroke survivors perform 19 tasks divided into 4 sub-groups \cite{mcdonnell2008action}. Task 1 to 16 involves reach and grasp and object actions followed by object transportation from point A to point B. Task 17 to 19 involves gross motor movement. The ARAT setup involves a chair and table, an activity mat (used as a starting position for the exercises and also for initial hand placement location), a wooden shelf (used as a target location for the object transportation exercises), and 15 different objects (such as wooden blocks, marbles, and drinking cups). The details about the tasks and objects can be found in \cite{mcdonnell2008action}. Clinicians follow a standardized protocol (site) to assess the performance of stroke survivors for these 19 tasks and provide a score between 0 to 3 for each task. During the exercises, the clinicians move around the patient in the space and sometimes lean over the patient. The correct finger position and orientation is an important factor in the scoring. We observed several ARAT sessions by clinicians and discussed with clinicians how to best instrument video capture of the ARAT so as a) to be unobtrusive, b) easily implemented by the clinicians while they also administer the ARAT and without technical help, and c) capture all pertinent visual data for assessment. The resulting setup included a camera array with four RGB cameras and a tablet interface that can be used to control the recording and administer the ARAT test. Three of the cameras were attached to the ARAT table and one on the side wall so as to capture the ipsilateral, contralateral, transverse, and back views of the activity space. The cameras capture RGB videos at a 4k resolution. Figure. \ref{fig:cap} demonstrates the capture setup in the clinical environment as well as the views captured by the 3 cameras on the table. 
\begin{figure}[h]
  \centering
  \includegraphics[width=\linewidth]{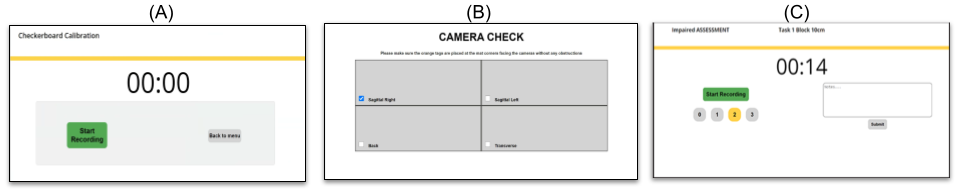}
  \caption{Calibration screen, camera check screen, and ARAT administration screen of our custom-developed capture interface}
  \label{fig:cap}
\end{figure}

The capture interface had three main screens: calibration screen, camera check screen, and ARAT administration screen. Before the first capture session of the day, the clinicians used our custom-developed capture interface to first calibrate the 4 RGB cameras using fiducial markers and a checkerboard \cite{kelliher2022calibrating}. With the cameras calibrated, the clinicians used the camera check page to confirm that the cameras were functioning and capturing the correct information. The clinicians then used the administration screen to start and stop the video recording at the beginning and end of the performance of each task by the stroke survivors thus producing separate videos for each task. The start-stop buttons also controlled the task timer which is a necessary component of the standardized assessment. After stopping the recording, the clinicians used the administration screen to enter a preliminary assessment for each task and note any problems in the execution and/or recording of the task.  

We conducted a pilot study where 5 clinicians administered the ARAT test to stroke survivors while also using the capture system. The final video results (from all cameras) were shown to the clinicians. We finished the discussion with a short survey shown in Table \ref{tab:rec}. Based on the clinician's feedback, we changed the viewing frame of the contralateral camera to have 2.5 times zoom-in on the patient's hand and the object so as to facilitate a detailed assessment of hand-object relation. We introduced the option of slow playback in both the segmentation and rating interfaces (see related sections below).  We finally used the data to establish preferred views for each segment: the viewing angle (and related camera view) that most therapists preferred for assessing each type of segment. However, we also noted that all other camera views need to be available to segmentors and raters since not all clinicians prefer the same view. Furthermore, for some segments, clinicians used at least two different views for assessment. 
\begin{table}[!h]
\centering
  \caption{Clinician Questionnaire at the End of Pilot Capture Study}
  \label{tab:rec}
\begin{tabular}{ll}
\toprule
 \textbf{Questionnaire} & \textbf{Answer} \\ 
\midrule
Did you find the Half Speed views helpful?                                                                                                                & 12 yes, 6 no                           \\ \hline
Which viewing angle did you find most helpful?                                                                                                            & 10 contralateral, 8 ipsilateral        \\\hline 
Did you find the Enhanced Termination view helpful?                                                                                                       & 10 yes, 8 No                           \\ \hline
\begin{tabular}[c]{@{}l@{}}Which viewing angle for the Enhanced \\ Termination did you find most helpful?\end{tabular}                                    & 10 Contralateral, 6 Ipsilateral        \\\hline 
\begin{tabular}[c]{@{}l@{}}Does the Enhanced Termination view assist you in \\ confidently determining a rating for the Termination segment?\end{tabular} & 11 yes, 7 No                           \\ 
\bottomrule
\end{tabular}
\end{table}
\subsection{ARAT Segmentation Vocabulary and Standardized Segmentation Tool}
In our prior work \cite{kelliher2020towards}, we introduced a four-component state machine with each state representing a type of movement segment used frequently in upper extremity rehabilitation. The state machine could be used to recreate all 15 upper extremity tasks used in our previous work. In this study, we modified the segments (states) as follows so they could be used, in different orders, to compose all 19 ARAT tasks: IP (initiation and progression), T (termination), MTR (manipulation and transport), PR (placement and release), GIP (gross initiation and progression) and GT (Gross termination). 
\begin{figure}[h]
  \centering
  \includegraphics[width=\linewidth]{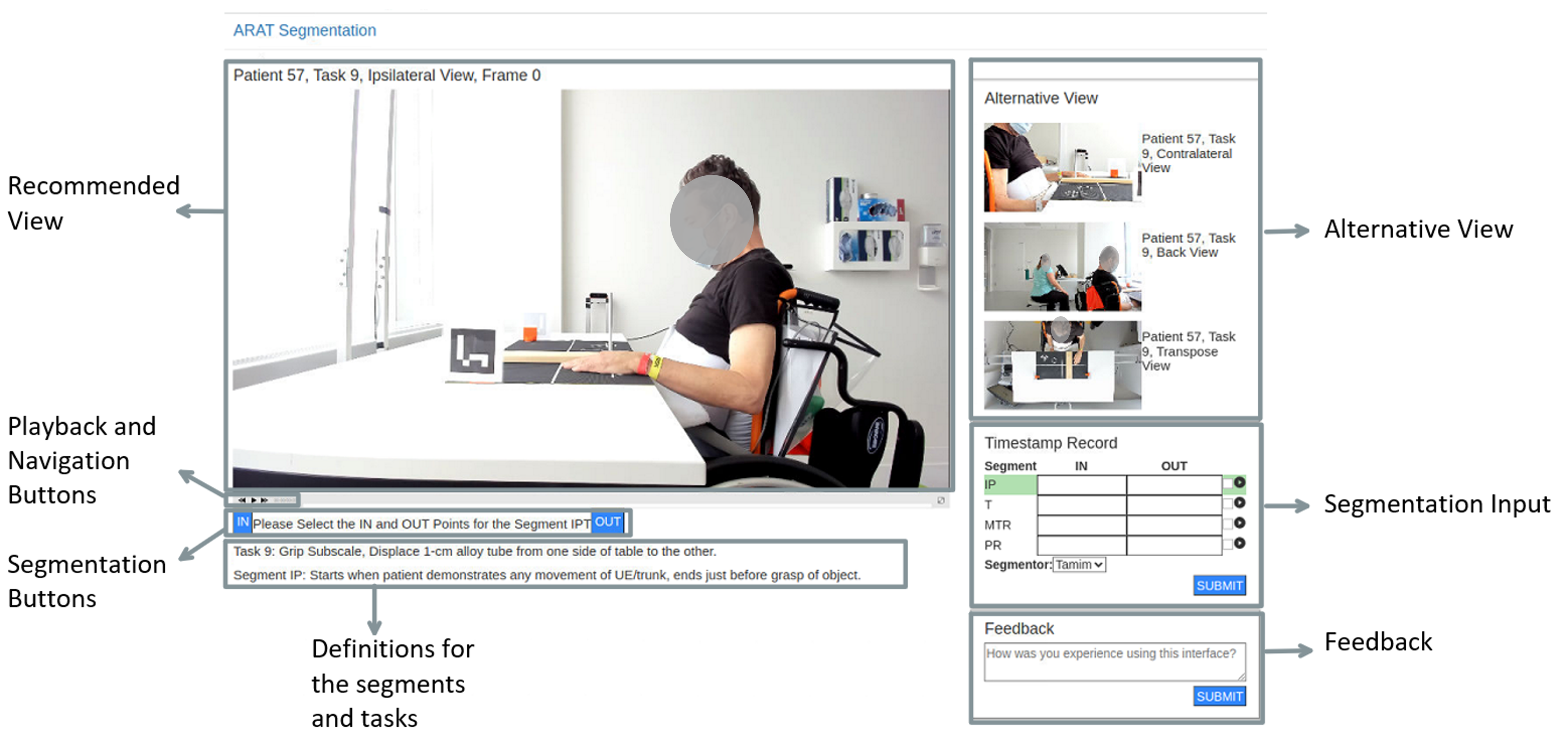}
  \caption{Segmentation interface and its key components}
  \label{fig:tool}
\end{figure}

To develop a segmentation tool that is compatible with clinical practice in the assessment of upper extremity rehabilitation, we designed a React-based interface as demonstrated in Fig. \ref{fig:tool}.  The interface loads a captured task from the clinic and retrieves (from the state machine) the expected sequence of segments. It then chooses the preferred view for the first segment and places the video from the preferred camera view in the main window. The interface then prompts the segmentor for an IN and OUT point for that segment. The "IN" and "OUT" buttons are right below the main screen. To make it easier for novice segmentors, the clinician developed definitions for the requested type of segments and that is also given below the video player. If the loaded viewing angle is insufficient for the segmentor, there is a list of alternative views of the same task on the right of the interface. Based on our preparatory work with the clinicians, we make sure the video player can play videos at half speed and frame by frame. This allows the segmentor to closely follow the clinical definitions for segmenting, For example, if a patient flexes his or her fingers before gripping an object, the segmentor can pause the video at the exact frame of the flexion and press the "IN" and "OUT" buttons to set this frame as the ending frame of the Initiation-Progression segment and the starting frame of the Termination segment. This process repeats for all the segments of a task and the recorder "IN" and "OUT" points for each segment are loaded in the segmentation table on the lower left of the screen.  

To prevent segmentation errors, we implemented various functionalities. First, segmentors can check each segmented "IN" and "OUT" points by clicking the playback button that opens a separate playback window that shows only the segment bounded by the selected "IN" and "OUT" points. Additionally, warning highlights will appear in the segmentation table if segments overlap, and segmentors must tick all the "IN OUT" boxes on the segmentation table before submission. Segmentors can also write feedback if there are any issues playing back the video or problems with the interface. With all these functionalities in one application, users can have a comprehensive understanding of patients' movements and create high-quality data with only simple clicks.  
\begin{table}[]
\centering
\caption{The structure of the Database Integrated with the Segmentation and Rating Tool}
\label{tab:data}
\begin{tabular}{lllllll}
\toprule
Patient Id & Hand & Task number & Segment name & Camera name   & Segment start frame & Segment end frame \\
\midrule
1          & left & 1           & IP           & Ipsilateral   & 75                  & 0                 \\
1          & left & 1           & IP           & Contralateral & 75                  & 92                \\
1          & left & 1           & T            & Contralateral & 92                  & 111    \\
\bottomrule
\end{tabular}
\end{table}

Based on the design of the segmentation interface we developed a relational database. The database stores critical information, like camera views, video filenames, segment definitions, user feedback, and segmentation data. The database is normalized to reduce redundancy, is applied to constraints to enforce data integrity, and is regularly monitored and maintained to ensure its reliability. To utilize the database, we make "GET" requests to display the videos, segments, and camera information in our interface, and make "PUSH" requests to the API whenever segmentors update a time entry or switch to different views. This database also underpins our segmentor behavior studies. For example, segmentor 1 starts interacting with the UI using the recommended Ipsilateral camera view, enters the segment start frame value of 75 for segment IP, and confirms it using the check box placed in the UI. We consider this as a unique step and keep that information. In the next step, the segmentor switches the camera by selecting a different camera view, in this case, the contralateral camera view, then updates the segment end frame to 92 and confirms it. Finally, the segmentor moves on to segment T, updates the segment start and end frame to 92 and 111 respectively, confirms it, and submits the form. Table \ref{tab:data} shows how we capture the information in a flattened tabular format considering the scenario described above. 

This format of the database allows us to directly call an API to share one common database with multiple services. The rating interface that facilitates the clinician to re-validate and rate the segmented videos uses the same database as well. Firstly, an API request is initiated to pull the "In" and "Out" points that represent the start frame and end frame of a segment, and according to the value, the respective video is trimmed and later displayed in the rating interface (see below). The clinician then reviews it, rates it, and finally submits it to the API which gets recorded in the same database.  The computational engine can now easily call the API to get access to the videos of tasks, segments, segmentation times, and the ratings associated with these. 
\begin{figure}[h]
  \centering
  \includegraphics[width=\linewidth]{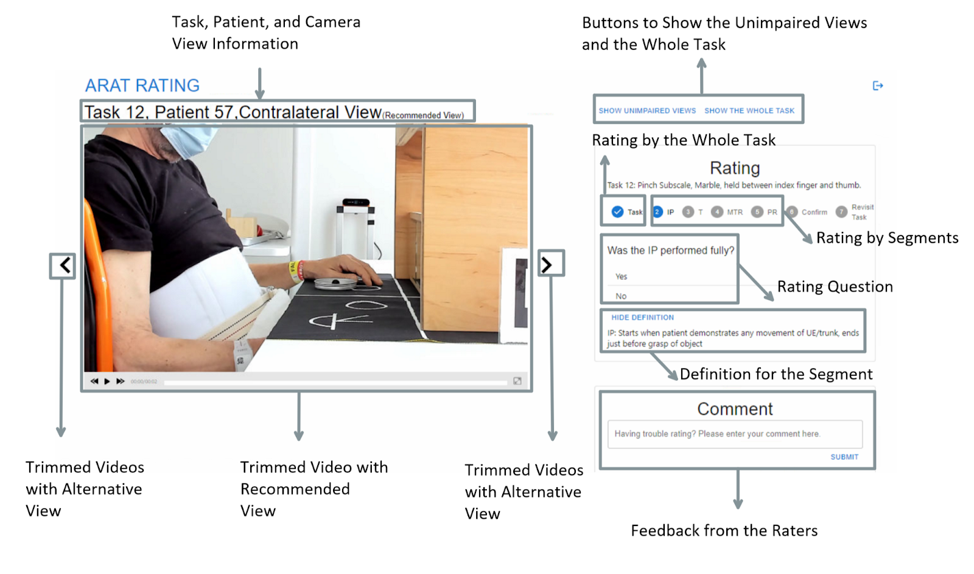}
  \caption{Rating interface and its key components}
  \label{fig:rat}
\end{figure}
\subsection{Rating Interface}
The rating interface displayed in Fig. \ref{fig:rat} is built upon the foundation of the segmentation interface.  The task and trimmed videos are retrieved from the database and displayed within the rating interface. Positioned on the left side of the interface are these videos with recommended view displayed by default, and arrow buttons enabling raters to view patient movements from different angles. On the right-hand side, a rating box prompts raters with questions regarding patients' performance, which facilitates the calculation of rating scores addressing task, segment, and component performance in an integrative manner.  To enhance rater understanding, we provide task and segment definitions, along with buttons offering unimpaired views of movements as reference. Additionally, to continuously improve the interface, we've incorporated a feedback box, which allows raters to communicate technical or rating-related concerns with us. (for more info on rating interface design see \cite{kelliher2020towards}).

\section{Results}
\subsection{Data Acquisition Summary}
With the multi-camera capture setup, we have captured 106 stroke survivors performing the ARAT in a major rehabilitation clinic. The total number of individual task videos in our dataset is 1800 and due to technical issues, only 50 videos are not usable. Using the segmentation tool, three segmentors (with little experience in rehabilitation practices) have segmented 760 task videos resulting in an average of 3000 segment videos. The clinicians have already rated 90\% of these segmented videos. 

\subsection{Segmentation Interface Usability}
\begin{figure}[h]
  \centering
  \includegraphics[width=\linewidth]{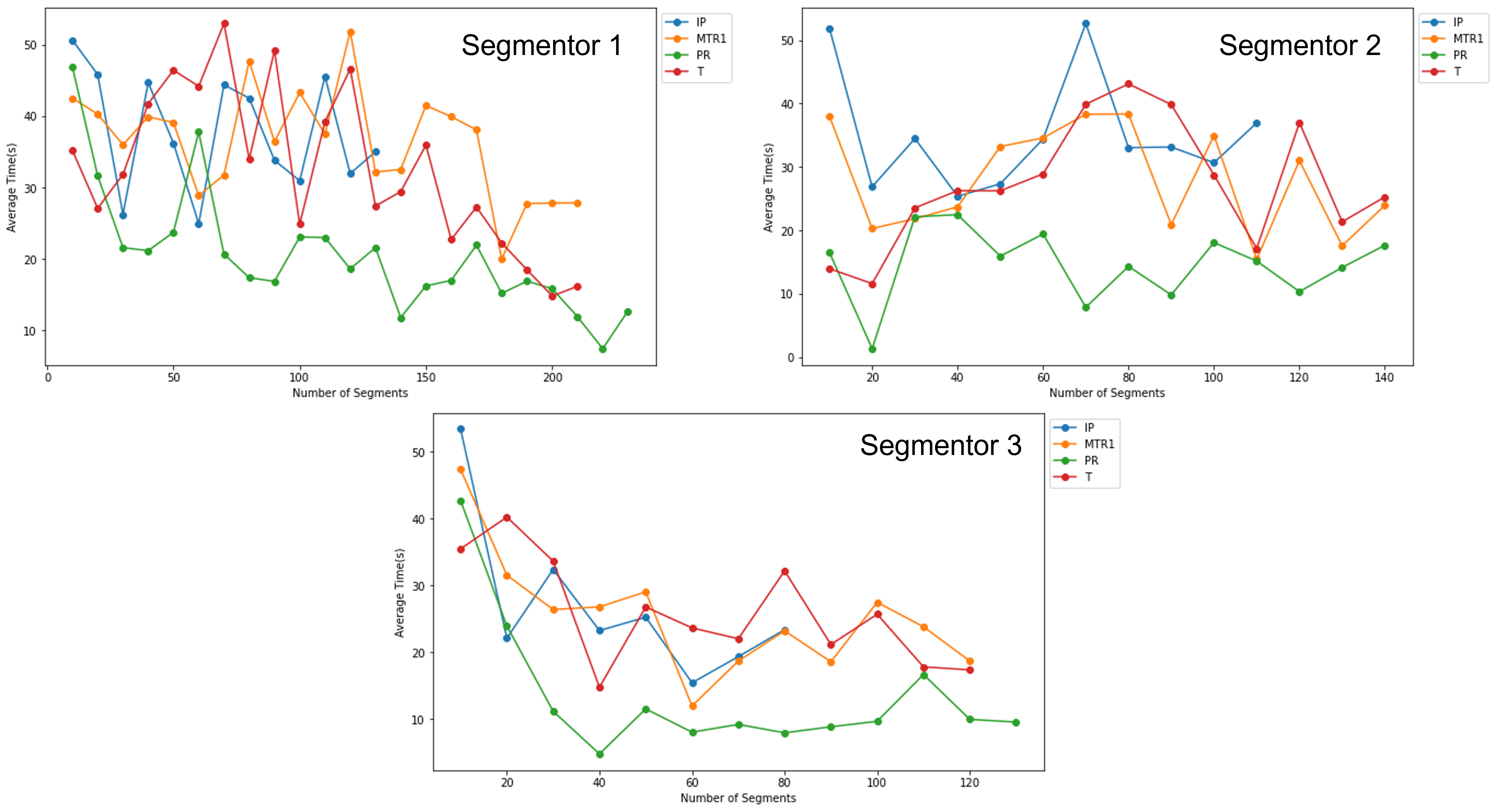}
  \caption{This plot demonstrates the decrease in segmentation time with the number of segmented videos for 3 segmentors as their experience with the interface increases. On the y-axis, we have the average time taken to complete 10 segments. Colored lines indicate different types of movement segments. }
  \label{fig:tool_seg}
\end{figure}

To study the users' behavior in our segmentation interface, we conducted data analysis in terms of completion time for a segment. To do the analysis, we first did data cleaning by removing data outliers, then we calculated the time duration for each segment. After we obtained the time duration, we grouped the data in a batch size of 10 and calculated the average completion at each batch. We demonstrate the segmentation time vs. the number of segment videos segmented in Fig. \ref{fig:tool_seg}. The outcomes of our analysis revealed a noteworthy trend: users generally exhibited decreasing completion time for segments and after segmenting approximately 30 videos of tasks convergence toward 20 seconds per segment. This suggests that the interface is easy to learn and helps the user quickly achieve improved efficiency in their work. 

We demonstrate the average percentage usage of the recommended view across three segmentors for different ARAT movement segments in Fig. \ref{fig:tool}. It is evident that for all segments, the recommended camera was the one mostly used by all three segmentors. However, for many segments (and especially for segments IP and MTR2), the segmentors preferred to switch to an alternative view. In Table. \ref{tab:switch}, we display the percentage of camera switching to input the starting and end frames per segment. As evident from the Table, there is a high switching occurrence during the transition to other segments. In IP and PR there is a comparatively high switching percentage for the starting frame since segmentors need to clearly visualise muscle twitching and release of the object that is indicative of the starting of IP and PR, respectively. In addition, to ensure accurate segmentation, the segmentors switch to an alternative camera and use the playback button to cross-check their provided segmentation times. This is also evident from the Table as there is a high percentage of cross-checking. Sometimes, during cross-checking, the segmentors would correct their prior entry. We see a high percentage of correction for the end frames since the segment transitions are critical for the clinicians. However, the overall, switching percentage is less than 25\% of the total entry meaning that the segmentors preferred to use the recommended view most of the time. These segmentors behaviors correlate highly with the clinician's preference for viewing angles per type of segment. This denotes that the segmentation interface facilitates the replication of tacit clinician segmentation practices by the novice segmentors.
\begin{figure}[h]
  \centering
  \includegraphics[width=0.6\linewidth]{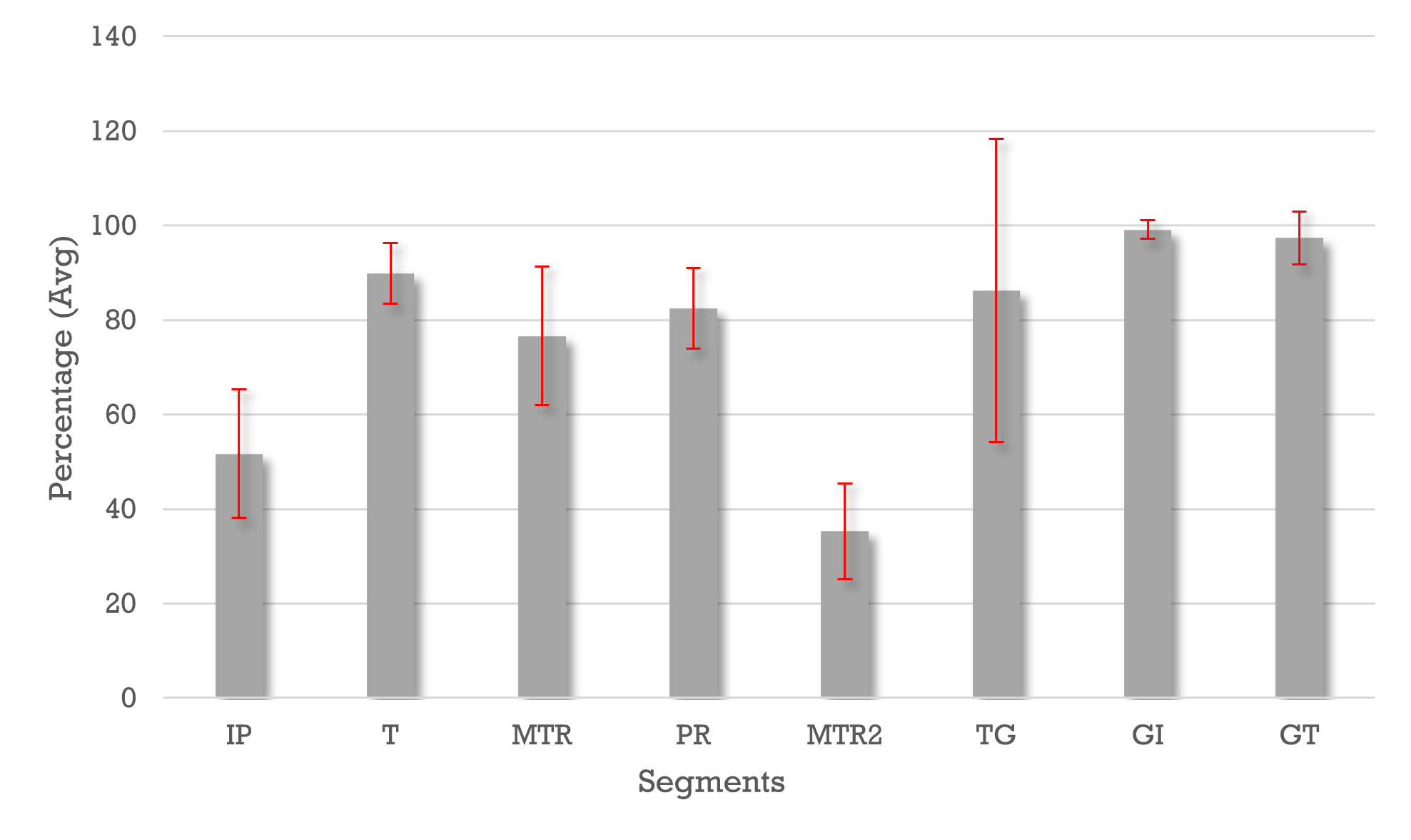}
  \caption{This plot demonstrates the average percentage $\pm$ standard deviation of using the recommended view per type of segment across three segmentors}
  \label{fig:tool}
\end{figure}
\begin{table}[!h]
\centering
\caption{The relationship between switching the camera views to the correction of start and end frames. All the values(\%) are calculated with respect to camera switching. }
\label{tab:switch}
\begin{tabular}{@{}lrrrrr@{}}
\toprule
\multicolumn{1}{c}{\begin{tabular}[c]{@{}c@{}}Segment\\ Names\end{tabular}} & \multicolumn{1}{c}{\begin{tabular}[c]{@{}c@{}}\% of Start \\ Frame Input\end{tabular}} & \multicolumn{1}{c}{\begin{tabular}[c]{@{}c@{}}\% of End \\ Frame Input\end{tabular}} & \multicolumn{1}{c}{\begin{tabular}[c]{@{}c@{}}\% of Start Frame \\ Correction\end{tabular}} & \multicolumn{1}{c}{\begin{tabular}[c]{@{}c@{}}\% of End Frame \\ Correction\end{tabular}} & \multicolumn{1}{c}{\begin{tabular}[c]{@{}c@{}}\% of Checking\\  Prior Input\end{tabular}} \\ \midrule
IP                                                                          & 15.5                                                                                   & 33.77                                                                                & 4.5                                                                                         & 13.14                                                                                     & 60.8                                                                                      \\
MTR                                                                         & 2.5                                                                                    & 30.6                                                                                 & 0.43                                                                                        & 15.5                                                                                      & 66.81                                                                                     \\
PR                                                                          & 13.13                                                                                  & 27.7                                                                                 & 0.73                                                                                        & 10.21                                                                                     & 67.88                                                                                     \\
T                                                                           & 6.33                                                                                   & 22.8                                                                                 & 2.53                                                                                        & 6.3                                                                                       & 73.42                                                                                     \\
MTR2                                                                        & 2.1                                                                                    & 36.1                                                                                 & 1.03                                                                                        & 14.4                                                                                      & 62.89                                                                                     \\ \bottomrule
\end{tabular}
\end{table}
\subsection{Rating summary}
Five Clinicians have rated 700 task videos using our custom-developed rating tool. They sought technical assistance from our team only for the first 5\% of the videos. The clinicians can provide feedback if the quality of the segmentation is not up to their standard. Some of the feedback from the clinicians were: "video does not actually show the initiation part" or "T video is too long". These indicate that there was an error in the segmentation. Based on their feedback, we corrected the segmentation times and made changes to the interface. However, we have received such feedback for only 4\% of the total tasks segmented and most of them were at the initial stage. This denotes that the segmentation results produced by novice segmentors are acceptable to the clinician and compatible with clinical practice.

\section{Discussion and Conclusion}
Explainable AI can lead to augmented intelligence practices in health care where the AI tools help support and advance the work of clinicians. Augmented intelligence in healthcare requires a design process that discovers and records the uniquely human aspects of the clinicians' expertise and develops AI processes that can intuitively synergize with the clinical practice.  In this paper, we present a participatory design methodology for acquiring data that can drive AI tools that produce recommendations that are compatible with clinical practice.  

Through collaboration with clinicians, we have developed a low-cost multi-camera capture system that was installed in a major rehabilitation clinic. The positioning and viewing parameters of the cameras were set to provide clinicians with all views necessary for rating the captured videos using the standardized ARAT protocol while addressing issues of functionality (task performance) and movement quality (segment and component performance) in an integrative manner. Five different clinicians were trained to calibrate the system and run the capture while also administering the ARAT test.  Training in the use of the system took less than 4 hours. Daily calibration took less than 5 minutes.  Sessions with 106 patients were captured successfully leading to 1800 individual videos with only 50 of these videos having some technical problems. The clinicians asked for technical assistance only five times across all 106 captures. Preliminary computational analysis of the videos shows that all key kinematic parameters needed for automated assessment can be extracted from the captured videos. We, therefore, propose that the capture system met its design goals.  It was unobtrusive and easy for clinicians to operate while also capturing all key data parameters for driving automated assessment. 

Expert clinicians worked closely with our team to establish a standardized segment vocabulary that could be used to segment all ARAT tasks and be compatible with clinician strategies in rehabilitation assessment. We established easily visible action characteristics (that involved combinations of limb location, object location, time, and speed) that could be used by novice segmentors for ARAT task segmentation. We also established the preferred camera view for each type of segment. We then developed an interface that assisted novice segmentors in segmenting the captured videos using the standardized segment vocabulary and produced results that are consistent with the way clinicians tacitly segment movement when assessing rehabilitation. We, therefore, expected that clinicians could easily use the segmentation results for an integrated rating of the captured tasks and their segments.  The segmented and rated videos could then be used to train machine learning algorithms for the automated segmentation and ratings of videos. 

The learning curve for all segmentors was short. Segmentors quickly converged to an average time of 20 seconds per segment.  Segmentors used the recommended camera for the majority of their segmentations and used multiple camera views for segments where clinicians also changed the point of view for complete assessment. The capture and segmentation results were well accepted by clinicians. 700 of the 760 segmented videos have been successfully rated by clinicians with each video receiving ratings from two different clinicians. The clinicians have reported segmentation problems with only 28 of the rated videos <4\%. In our prior work \cite{ahmed2021automated}, (without a standardized segmentation interface) clinicians reported segmentation inaccuracies for 10\% of the videos. We, therefore, propose that the segmentation interface allowed novice segmentors to quickly learn and then simulate the segmentation process of the clinicians. 

The presented process in this paper can be generalized into a participatory design methodology for acquiring data for augmented intelligence tools in many situations that involve expert assessment of complex human activity (health, sports, art performance, surgery, military, and first respondent training). The key components of the methodology are:
- Observe the assessment process by experts and record it on video.
- Ask experts to describe their process and then ask them to repeat it, step by step, while assessing videos of performances that cover the expected range of performance variability
- Develop capture tools that are unobtrusive, can be easily used by experts while they are also involved in their daily practice, and focus on capturing only the key components of assessment used by experts.
- Develop a consensus activity segmentation vocabulary (and related process) that captures the explicit and/or tacit segmentation approaches of multiple experts
- Develop a segmentation tool that allows novice segmentors to simulate the expert segmentation processes
- Develop a consensus rating process and rating tool that allows different experts to rate the totality of the action and its key components in an integrative and standardized manner.
- Use the resulting data for training AI tools that produce results compatible with expert practice and can thus contribute to augmented intelligence.

\bibliographystyle{ACM-Reference-Format}
\bibliography{sample-base}


\end{document}